\documentclass[11pt]{4ssmmp} 

\usepackage{amsfonts}
\usepackage{fancyhdr}
\usepackage{graphicx}
\usepackage{rotating}

\begin{document}
 \baselineskip=11pt

\title{New insights into abelian topologically
massive gauge theories\hspace{.25mm}\thanks{\,Work supported by the ``Fonds pour la formation \`a la Recherche dans l'Industrie et dans l'Agriculture'' (FRIA), Associated Funds of the FNRS, Belgium.}}
\author{\bf{Bruno Bertrand}\hspace{.25mm}\thanks{\,e-mail address:
bruno.bertrand@fynu.ucl.ac.be}
\\ \normalsize{Institut de Physique Nucl\'eaire and} \\ \normalsize{Center for Particle Physics and Phenomenology (CP3)} \\ \normalsize{Universit\'e catholique de Louvain} \\ \normalsize{Chemin du Cyclotron 2}\\ \normalsize{B-1348 Louvain-la-Neuve, Belgium}}

\date{}

\maketitle

\begin{abstract}

Abelian topologically massive gauge theories (TMGT) provide a topological mechanism to generate mass for any $p$-tensor boson in any dimension. Within the Hamiltonian formulation, the embedded topological field theory (TFT) is not made manifest. We therefore introduce a gauge invariant factorization of the classical phase space in two orthogonal sectors. The first of these sectors is that of gauge invariant dynamical variables with massive excitations. The second is that of a decoupled TFT. Through canonical quantization, a factorization of quantum states arises, enabling the projection from TMGT onto topological quantum field theories in a most transparent way.

\end{abstract}

\newcommand{\ud}{\mathrm{d}}

\section{Topologically massive gauge theories in any dimension}

Phase transitions are often associated with spontaneous symmetry breaking, for example in the BCS theory of superconductivity where condensation of Cooper pairs arises below the critical temperature. Furthermore, the symmetry breaking mechanism provides masses for the weakly interacting gauge vector bosons in the Standard Model, while remaining consistent with the renormalizability and unitarity constraints. However, the predicted Higgs boson has yet to be discovered. Within this context, the quest for alternative mechanisms for generating mass is quite fascinating.

Analogies between theoretical particle physics and condensed matter may again be fruitful. 
For example, Chern-Simons terms account for fer\-mio\-nic collective phenomena, such as in the quantum Hall effect. Likewise, the statistical transmutation of extended objects in higher dimensions relies on topological couplings of the ``$B\wedge F$'' type. When considered alone these types of terms are characteristic of topological quantum field theories (for a review see \cite{Birmingham:1991ty}). They actually possess so large a gauge freedom that their physical (gauge invariant) observables depend solely on the diffeomorphism equivalence class of the underlying manifold. 

Within the context of particle physics, masses for tensor bosons may originate from a topological mechanism preserving exact gauge invariance. Consider a real-valued $p$-form field $A$ in $\Omega^p(\mathcal{M})$ and a real-valued $(d-p)$-form field $B$ in $\Omega^{d-p}(\mathcal{M})$ over a ($d+1$) dimensional manifold $\mathcal{M}$. The general action of these ``topologically massive gauge theories'' reads
\begin{eqnarray} \label{def:TMGT_Action}
S[A,B] & = & \int_{\mathcal{M}} \frac{\sigma^{p}}{2 \, e^2}\, F\wedge\ast F + \frac{\sigma^{d-p}}{2 \, g^2} \, H\wedge\ast H \nonumber\\ 
& & + \kappa \int_{\mathcal{M}} (1-\xi) \, F\wedge B - \sigma^p \, \xi \, A\wedge H \, ,
\end{eqnarray}
where $\sigma = (-1)$. This action is invariant under two independent types of gauge transformations given by
\begin{eqnarray} \label{def:BF_gauge}
\delta_{\alpha} A = \alpha, &\quad& \delta_{\alpha} B = 0; \nonumber\\
\delta_{\beta} A = 0, &\quad& \delta_{\beta} B = \beta  ,
\end{eqnarray}
where $\alpha$ and $\beta$ are, respectively, $p$ and $(d-p)$-closed forms while $F=\ud A$ and $H=\ud B$ are the gauge invariant field strengths of $A$ and $B$, respectively. The scaling parameters $e$ and $g$ are real while $\kappa$ is a multiplicative constant. The arbitrary real variable $\xi$ parametrizing the surface term is irrelevant for appropriate boundary conditions. The action (\ref{def:TMGT_Action}) includes two dynamical $p$- and $(d-p)$-form fields $A$ and $B$ coupled through topological mass terms of the ``$B\wedge F$'' type. Through an appropriate gauge choice, it is possible to make one of the tensor fields massive by absorbing the physical degrees of freedom of the other tensor field. In 3+1 dimensions, one recovers the famous Cremmer-Scherk action \cite{TMGT_3+1D}. In the particular case where $d=2p$ with $p$ odd, a topological Chern-Simons term generates a mass for a propagating $p$-form field $A$,
\begin{equation} \label{def:MCS_Action}
S[A] = \int_{\mathcal{M}} \frac{-1}{2} \, F\wedge\ast F + \frac{\kappa}{2} \, A\wedge F .
\end{equation}
In 2+1 dimensions, one recognizes the Maxwell-Chern-Simons theory \cite{MCS_2+1D}.

\section{Factorization of the classical Hamiltonian}

In order to develop the Hamiltonian formulation of the above dynamics, the manifold $\mathcal{M}$ is chosen to have the topology $\mathcal{M}=\mathbb{R} \times \Sigma$ where $\Sigma$ is a compact and orientable $d$-dimensional space manifold without boundary. We may then adopt synchronous coordinates on $\mathcal{M}$, for which the metric takes the form\footnote{Latin indices $i = 1,\ldots,d$ label spatial directions in $\Sigma$.}
\begin{displaymath}
\ud s^2 = \ud t^2 - \tilde{h}_{ij}\, \ud x^i\, \ud x^j 
\end{displaymath}

Consider now a canonical split of the $p$-form field $A$, $A = \ud t \wedge A_0 + \tilde{A}$. A similar decomposition applies to the $(d-p)$-form $B$. The phase space variables are then restricted to elements in $\Omega^p(\Sigma)$ and $\Omega^{(d-p)}(\Sigma)$, $\tilde{A}$ and $\tilde{B}$, along with their conjugate momenta $\tilde{P}$ and $\tilde{Q}$, respectively. These variables possess the non vanishing Poisson Brackets
\begin{eqnarray}
   \left\lbrace A_{i_1\cdots i_p}(\vec{x},t) , P^{i_1\cdots j_p}(\vec{y},t) \right\rbrace &=&  \delta_{\left[ i_1\right. }^{j_1} \ldots \delta_{\left. i_p\right]}^{j_p} \, \delta(\vec{x} - \vec{y}), \nonumber\\
   \left\lbrace B_{i_1\cdots i_{d-p}}(\vec{x},t) , Q^{j_1\cdots j_{d-p}}(\vec{y},t) \right\rbrace &=&  \delta_{\left[ i_1\right. }^{j_1} \ldots \delta_{\left. i_{d-p}\right] }^{j_{d-p}} \, \delta(\vec{x} - \vec{y}) \nonumber\, .
\end{eqnarray}
The time components of the fields along with their conjugate momenta ($A_0$ and $P^0$, $B_0$ and $Q^0$, respectively) are auxiliary fields which are usually reduced from the Hamiltonian first-order action to lead to the fundamental Hamiltonian formulation of the system. The latter results
from the analysis of constraints \cite{Ham.quant.} of (\ref{def:TMGT_Action}) and reads,
\begin{eqnarray}\label{def:TMGT_Hamiltonian}
   H & = & \frac{e^2}{2}\, \left(\ast\tilde{P}-\kappa\, (1-\xi)\, \tilde{B}\right)^2 +  \frac{(\ud \tilde{A})^2}{2\, e^2} \\
     & + & \frac{g^2}{2}\, \left(\ast \tilde{Q}+\kappa\, \xi\, \sigma^{p(d-p)}\tilde{A}\right)^2 + \frac{(\ud \tilde{B})^2}{2\, g^2} + (\textrm{Surface term}) \nonumber\\
& + & \int_{\Sigma} \sigma^p\, u' \wedge\ud\left(\ast\tilde{P}+\kappa\, \xi\, \tilde{B}\right) + \sigma^{d-p}\, v'\wedge\ud\left(\ast\tilde{Q}-\kappa (1-\xi) \sigma^{p(d-p)}\, \tilde{A}\right), \nonumber
\end{eqnarray}
where henceforth one assumes that the Hodge $\ast$ operation is restricted to the space submanifold $\Sigma$. In (\ref{def:TMGT_Hamiltonian}) the inner product on $\Omega^k(\Sigma)\times\Omega^k(\Sigma)$ is defined as
\begin{equation}\label{def:inner_prod}
    (\omega_k)^2 = \left(\omega_k,\omega_k\right) \quad \textrm{with}\quad \left(\omega_k,\eta_k\right) = \int_{\Sigma} \omega_k\wedge\ast\eta_k  .
\end{equation}
Finally in (\ref{def:TMGT_Hamiltonian}), $u'$ and $v'$ are arbitrary Lagrange multipliers for the two ``Gauss law'' first-class constraints which generate the two types of gauge transformations in (\ref{def:BF_gauge}) which are continuously connected to the identity transformation, namely the so-called ``small gauge transformations''.

Within the Hamiltonian formulation, the topological terms in (\ref{def:TMGT_Action}) and (\ref{def:MCS_Action}) are responsible both for non vanishing brackets between the ``electric fields'' and for generating a mass gap. However, the identification of a TFT sector within a TMGT is not manifesta and the basic variables used to define the theory do not create physical states since they are not gauge invariant degrees of freedom. We therefore introduce a factorization of the classical phase space in two orthogonal sectors\footnote{Poisson Brackets between variables from the two distinct sectors vanish.}, while requiring also that the transformation preserves the canonical commutation relations. This factorization replaces usual gauge fixing procedures (and the ensuing introduction of second-class constraints) with a canonical transformation in order to isolate already at the classical level physical variables from gauge variant degrees of freedom \cite{Bertrand:2007(1)}.

Let us first introduce the following gauge variant variables,
\begin{equation}\label{def:Tsec_variables}
   \mathcal{A} = -\frac{1}{\kappa} \, \sigma^{p(d-p)} \, \ast\tilde{Q}+(1-\xi)\, \tilde{A} \quad , \quad \mathcal{B} = \frac{1}{\kappa}\ast\tilde{P}+\xi\, \tilde{B} \, ,
\end{equation}
defined on the dual sets $\Omega^p(\Sigma)$ and $\Omega^{d-p}(\Sigma)$.
They are canonically conjugated,
\begin{displaymath}
   \left\lbrace \mathcal{A}_{i_1\cdots i_p}(\vec{x},t) , \mathcal{B}_{j_1\cdots j_{d-p}}(\vec{y},t) \right\rbrace = \frac{1}{\kappa}\, \epsilon_{i_1\cdots i_p j_1\cdots j_{d-p}} \, \delta(\vec{x}-\vec{y})  ,
\end{displaymath}
and transform under the two types of gauge transformations (\ref{def:BF_gauge}) as
\begin{eqnarray} \label{eq:Tsec_gauge}
\delta_{\alpha} \mathcal{A} = \alpha, &\quad& \delta_{\alpha} \mathcal{B} = 0; \qquad \nonumber\\
\delta_{\beta} \mathcal{A} = 0, &\quad& \delta_{\beta} \mathcal{B} = \beta  .
\end{eqnarray}
The remaining independent and conjugated physical degrees of freedom are
\begin{equation}\label{def:Psec_variables}
   G = \tilde{Q}+\kappa\, \xi\, \ast\tilde{A} \quad , \quad 
   E = \tilde{P}-\kappa\, (1-\xi)\, \sigma^{p(d-p)}\, \ast\tilde{B} ,
\end{equation}
\begin{displaymath}
   \left\lbrace E^{i_1\cdots i_p}(\vec{x},t) , G^{j_1\cdots j_{d-p}}(\vec{y},t) \right\rbrace = -\kappa \, \epsilon^{i_1\cdots i_p j_1\cdots j_{d-p}} \, \delta(\vec{x}-\vec{y}) .
\end{displaymath}
Based on the equations of motion, it may be seen that these variables correspond to the non commutative ``electric'' fields associated to $A$ and $B$, respectively. As stated earlier, these two sectors are orthogonal.

Finally, the Lagrange multipliers in (\ref{def:TMGT_Hamiltonian}) may be specified in a manner convenient to obtain the factorized fundamental Hamiltonian,
\begin{eqnarray}\label{def:Psec_Ham}
   H[E,G,\mathcal{A},\mathcal{B}] & = & \frac{e^2}{2}\, \left(E\right)^2 + \frac{(\ud^\dagger E)^2}{2\, \kappa^2\, g^2} + \frac{g^2}{2}\, \left(G\right)^2 + \frac{(\ud^\dagger G)^2}{2\, e^2 \, \kappa^2} \nonumber\\
   & + & \kappa\, \int_{\Sigma} \sigma^p\, \mathcal{A}_0\wedge\ud\mathcal{B} - \sigma^{(p+1)(d-p)}\, \mathcal{B}_0\wedge\ud\mathcal{A} \, , 
\end{eqnarray}
where $\ud^\dagger = *d*$ is the coderivative operator. Obviously, $\mathcal{A}_0$ and $\mathcal{B}_0$ are Lagrange multipliers enforcing
the first-class constraints which generate the small gauge transformations of (\ref{eq:Tsec_gauge}).
The first sector (\ref{def:Tsec_variables}) is that of non propagating variables identified to define precisely a TFT of Chern-Simons or ``$B\wedge F$'' type; the ``TFT'' sector. The second sector of gauge invariant variables (\ref{def:Psec_variables}) characterizes massive dynamical degrees of freedom.
\begin{eqnarray*}
   \begin{array}{c}
 \textbf{Original} \\
 \textbf{phase space}
\end{array}
 \left\{ \begin{array}{|c|}
 \hline
\tilde{A}\, , \tilde{P} \\
\hline\hline
\tilde{B}\, , \tilde{Q} \\
\hline
\end{array}
 \right.&& \\
\centering{\Downarrow} & \textit{Factorization} & \\ &&\\
\begin{array}{c}
 \textbf{Factorized} \\ \textbf{phase space}
\end{array}
\left\{
\begin{array}{|c|}
 \hline
\mathcal{A}\, , \mathcal{B} \\
\hline\hline
E \, , G \\
\hline
\end{array}
\right. &
\begin{array}{c}
 \textrm{``TFT'' sector} \\
 \textrm{Dynamical sect.}
\end{array}
&
\begin{array}{l}
 \} \textrm{Gauge variant variables} \\
 \} \textrm{Physical variables}
\end{array}
 \end{eqnarray*}

Using Hodge duality between $\Omega^p(\Sigma)$ and $\Omega^{d-p}(\Sigma)$, one identifies the Hamiltonian of a $p$-form field of mass $m=\hbar\mu$,
\begin{displaymath}
H[C,E,\mathcal{A},\mathcal{B}] = \frac{\mu^2}{2}\, \left(C\right)^2 + \frac{1}{2}\, \left(\ud C \right)^2 + \frac{1}{2}\, \left(E\right)^2 + \frac{(\ud E)^2}{2\, \mu^2} + H_{\textrm{TFT}}[\mathcal{A},\mathcal{B}] ,
\end{displaymath}
given the following associations made with (\ref{def:Psec_Ham}),
   $\mu=\kappa\, e\, g$, $E \to \frac{E}{e}$ and $\ast G = e\, \kappa\, (-1)^{p(d-p)}\, C$, while
\begin{displaymath}
\left\lbrace C_{i_1\cdots i_p}(\vec{x},t) , E^{i_1\cdots j_p}(\vec{y},t) \right\rbrace = \delta_{\left[ i_1\right. }^{j_1} \ldots \delta_{\left. i_p\right]}^{j_p} \, \delta(\vec{x} - \vec{y}) \, .
\end{displaymath}
Note that the identification of the physical content of these theories in terms of a massive $p$-form (or ($d-p$)-form, by duality)
has been achieved without applying any gauge fixing procedure whatsoever, in contradistinction to all results available so far in the literature.

\section{Factorization of quantum states}

The new Topological-Physical (TP) factorization applies within the Hamiltonian formulation and circumvents gauge fixing. Dirac's quantization is readily implemented for small gauge transformations, and may also deal with large gauge transformations on homologically non trivial manifolds.
The quantized system is defined by a Hilbert space representation of the following commutation relations for the linear self-adjoint operators associated to the classical variables,
\begin{eqnarray}
\left[\hat{\mathcal{A}}_{i_1\cdots i_p}(\vec{x}) , \hat{\mathcal{B}}_{j_1\cdots j_{d-p}}(\vec{y}) \right] & = & \frac{i\,\hbar}{\kappa}\, \epsilon_{i_1\cdots i_p j_1\cdots j_{d-p}}\, \delta(\vec{x}-\vec{y}), \nonumber\\
\left[ \hat{E}^{i_1\cdots i_p}(\vec{x}) , \hat{G}^{j_1\cdots j_{d-p}}(\vec{y}) \right] &=& -\frac{i\,\hbar}{\kappa}\, \epsilon^{i_1\cdots i_p j_1\cdots j_{d-p}} \, \delta(\vec{x}-\vec{y}) \nonumber .
\end{eqnarray}
Hilbert space consists of functionals $\Psi[\mathcal{A},E]$. TP factorization being consistent with quantization \cite{Bertrand:2007(2)}, these wave functionals factorize into components associated to each sector of variables, 
\begin{displaymath}
   \Psi[\mathcal{A},E] = \Phi[E] \, \Psi[\mathcal{A}]  .
\end{displaymath}
Furthermore, based on the Hodge theorem each sector may be divided into three subsectors, as applies already at the classical level,
\begin{equation}\label{def:Hodge}
   \mathcal{A} = A^e + A^c + A_h, \qquad \mathcal{B} = B^e + B^c + B_h  ,
\end{equation}
each of these contributions being associated to an exact form, a coexact one (local part) and a harmonic part (global part), with respect to the inner product defined in (\ref{def:inner_prod}). The global part is in fact of a topological character\footnote{The harmonic part of a $p$-form is dual to the $p$-cohomology group $H^p(\Sigma,\mathbb{R})$.} and decomposes into topological invariants in a convenient basis,
\begin{displaymath}
   a_\gamma = \oint\limits_{\Sigma_{(p)}^\gamma} A_h \quad , \quad b_\gamma = \oint\limits_{\Sigma_{(d-p)}^\gamma} B_h.
\end{displaymath}
These quantities are simply the generalized Wilson loops for the generators $\{\Sigma_{(p)}^\gamma\}_{\gamma=1}^{\beta_p}$ and $\{\Sigma_{(d-p)}^\gamma\}$ of the dual homology groups $H_p(\Sigma)$ and $H_{d-p}(\Sigma)$ of rank $\beta_p$.
Commutators of these global operators are topological invariants,
\begin{displaymath}
\left[ a^{\gamma} , b^{\gamma'} \right] = i \frac{\hbar}{\kappa} I^{\gamma\gamma'} \quad , \quad 
I^{\gamma\gamma'} = \cap \left[\Sigma_{(p)}^\gamma,\Sigma_{(d-p)}^{\gamma'}\right] ,
\end{displaymath}
namely the elements of the intersection matrix each of which entries is the sum of signed intersections of the generators of $H_p(\Sigma)$ and $H_{d-p}(\Sigma)$.

The lack of dynamics for the gauge variant TFT part implies that $\Psi[\mathcal{A}]$ does not contribute to the energy spectrum. The gauge invariant physical wave functional $\Psi^p$ in that sector is that of a topological quantum field theory \cite{Szabo:1999gm}. Its invariance under small gauge transformations, according to Dirac quantization, requires it to belong to the kernel of the first-class constraint operators in (\ref{def:Psec_Ham}). This means that the ``TFT'' part of the physical wave functional depends only on global variables, $\Psi^p[a^\gamma]$. The degeneracy is further constrained by requiring invariance of physical states under large gauge transformations (LGT). This implies that $\kappa$ is quantized in terms of topological invariants,
\begin{equation}\label{eq:kappa_quant}
   \kappa = \frac{\hbar}{2\pi}\, \mathcal{I}\, k \quad , \quad k \in \mathbb{Z} \quad , \quad \mathcal{I} = \det\left( I^{\gamma\gamma'} \right) \in \mathbb{N}.
\end{equation}
In a weaker sense, requiring invariance of the Hilbert space under LGT implies that $\kappa=k_1/k_2$ be rational while the physical wave functional carries a $(k_2)^p$-dimensional projective representation of the algebra of LGT \cite{Szabo:1999gm}.

The dynamical component $\Phi[E]$ is already gauge invariant and contributes to the energy spectrum.
On the $d$-torus, diagonalization of the quantum Hamiltonian leads to the spectrum
\begin{equation}\label{eq:Total_Energy}
\varepsilon_{\left(n^\gamma(\underline{k})\right)} = \varepsilon_{(0)} + \hbar \, \sum_{\underline{k}\in\mathbb{Z}^d} \sum_{\gamma} n^\gamma(\underline{k})\, \sqrt{4\, \pi^2\, \omega(\underline{k})^2 + \mu^2}  ,
\end{equation}
where $\omega^2(\underline{k}) = k_i\, k_j\, \delta^{ij}$. This expression combines contributions from both the global and the local subsectors in the dynamical sector, see (\ref{def:Hodge}). Hence the index $\gamma$ has different meanings whether $\underline{k}\neq \underline{0}$ or $\underline{k}=\underline{0}$. In the former situation, $\gamma$ denotes polarization while in the latter it is a cohomology indice\footnote{On the $d$-torus, the $p^{\textrm{th}}$ Betti number, $\beta_p$, is equal to $C^p_d$. Thus $\gamma \in \{1,\ldots,C^p_d \}$.}.
The vacuum energy $\varepsilon_{(0)}$ diverges,
\begin{displaymath}
\varepsilon_{(0)} = \frac{\hbar}{2}\, C^p_d\, \sum_{\underline{k}\in\mathbb{Z}^d} \sqrt{4\, \pi^2\, \omega(\underline{k})^2 + \mu^2}  .
\end{displaymath}
The integer-valued functions $n^\gamma(\underline{k})$ count, for $\underline{k} \neq 0$, the number of massive quanta of a $p$ or $(d-p)$-tensor field of momentum $2\, \pi\, \hbar\, \underline{k}$ and rest mass
\begin{equation}\label{def:Mass_gap}
   M = \hbar\, \mu = \hbar\, \kappa\, e\, g \, ,
\end{equation}
which is the mass gap of this quantum field theory. There is also the contribution of global quantum states where $n^\gamma(\underline{0})$ count the number of excitations along the homology cycle generators.
For what concerns LGT, if $\kappa$ is a rational number of the form $\kappa=k_1/k_2$ each energy eigenstate is
$(k_1\, k_2)^{\beta_p}\, \mathcal{J}_p$-fold degenerate, a multiplicity of topological origin with
\begin{displaymath}
   \mathcal{J}_p = \prod_{\delta=1}^{\beta_p} \mathcal{I}\, \textrm{Min}(I_{\delta\delta'}) \quad \textrm{and} \quad
\sum^{N_p}_{\delta=1} I_{\gamma\delta}\, I^{\delta\gamma'} = \delta_\gamma^{\gamma'}.
\end{displaymath}
If $\kappa \in \mathbb{Z}$, each energy eigenstate is $k^{\beta_p}\, \mathcal{J}_p$-fold degenerate and the mass gap (\ref{def:Mass_gap}) takes only discrete values, see (\ref{eq:kappa_quant}).

Interestingly, the new TP factorization enables the definition of the usual projection\footnote{This projection is analogous to the lowest Landau level projection.} of TMGT onto TQFT in a natural way. As a matter of fact, the non commuting sector of a CS theory or, more generally, the reduced phase space of a TQFT appears no longer only after the projection onto the quantum ground state but is already manifest even at the classical Hamiltonian level. Letting $e$ or $g$ run to infinity, dynamical excitations  become infinitely massive, see (\ref{def:Mass_gap}), and in some sense the system looses its dynamics, the latter being intimately related to any dependence on the spacetime metric structure of the theory. All that is then left is thus a wave functional dependent on global topological variables only, associated to a TQFT, and independent of ``coupling constants''.

\section{Conclusion and future developments}

As discussed in separate work \cite{Bertrand:2007(2)}, the above TP factorized parametrization, (\ref{def:Tsec_variables}) and (\ref{def:Psec_variables}), can be extended to the covariant Lagrangian formulation, hence leading to the following general and completed structure,\\ \\
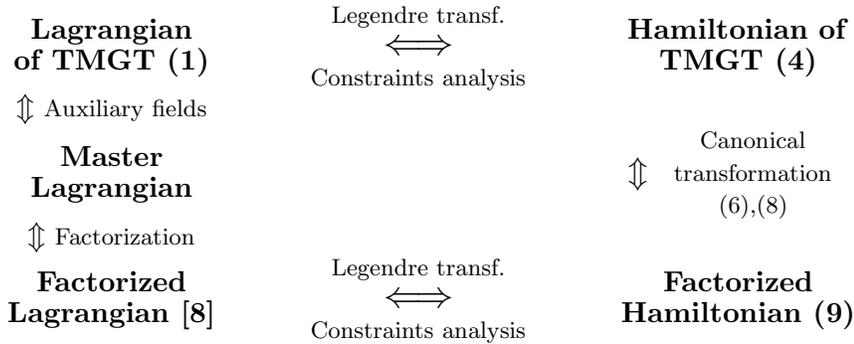
\begin{tabular}{ccccc}
   \begin{tabular}{c}
   \textbf{Lagrangian} \\ \textbf{of TMGT (\ref{def:TMGT_Action})}
   \end{tabular}
& & 
   \begin{tabular}{c}
   \footnotesize{Legendre transf.} \\ \Large{$\Longleftrightarrow$} \\ \footnotesize{Constraints analysis}
   \end{tabular}
& & 
   \begin{tabular}{c}
   \textbf{Hamiltonian of} \\ \textbf{TMGT (\ref{def:TMGT_Hamiltonian})}
   \end{tabular}
\\
$\Updownarrow$
   \footnotesize{Auxiliary fields}
& & \\
   \begin{tabular}{c}
   \textbf{Master} \\ \textbf{Lagrangian}
   \end{tabular}
& & & &
\large{$\Updownarrow$}
   \begin{tabular}{c}
   \footnotesize{Canonical} \\ \footnotesize{transformation} \\ \footnotesize{(\ref{def:Tsec_variables}),(\ref{def:Psec_variables})}
   \end{tabular}
\\
$\Updownarrow$
\footnotesize{Factorization} &&&&\\
   \begin{tabular}{c}
   \textbf{Factorized} \\ \textbf{Lagrangian \cite{Bertrand:2007(2)}}
   \end{tabular}
& & 
   \begin{tabular}{c}
   \footnotesize{Legendre transf.} \\ \Large{$\Longleftrightarrow$} \\ \footnotesize{Constraints analysis}
   \end{tabular}
& & 
   \begin{tabular}{c}
   \textbf{Factorized} \\ \textbf{Hamiltonian (\ref{def:Psec_Ham})}
   \end{tabular}
\end{tabular}\\
There exists a well-known duality between the Maxwell-Chern-Simons theory \cite{MCS_2+1D} and the ``self-dual'' massive model in 2+1 dimensions \cite{Deser:1984}. This dual description has been extended to topologically massive gauge theories in any dimension \cite{Menezes-Cantcheff}. The Lorentz covariant extension of the Hamiltonian PT factorization (\ref{def:Tsec_variables}) and (\ref{def:Psec_variables}) discussed here enables the construction of this type of duality by a change of variables in the ``master'' first-order action. In contradistinction to results of previous works, the dual action thereby obtained possesses the same gauge structure as the original theory. Furthermore, the dual action is factorized into a propagating sector of massive gauge invariant variables and a sector with gauge variant variables defining a topological field theory. Using a different approach, we recover in a more general context some of the results in \cite{Arias-Guimaraes}.

Based on the methods outlined in this contribution, we have also been able to show that in the symmetry breaking phase the effective abelian Maxwell-Higgs Lagrangian is equivalent to a particular form of TMGT coupled to a real scalar field in a very specific way~\cite{Bertrand:2007prep}. We hope that in the near future, it will become possible to develop new types of perturbation techniques when TMGT are coupled to matter fields. The basic idea is to determine the nonperturbative configurations resulting from the coupling to the ``TFT'' sector and then develop perturbatively the dynamical sector. 

\section*{Acknowledgement}

I would like to thank my supervisor, Prof. Jan Govaerts, for his contributions to this work and for his critical reading of the present note.

\end{document}